# Separating vascular and neuronal effects of age on fMRI BOLD signals

Kamen A. Tsvetanov[1,2,*], Richard N.A. Henson[3,4] and James B. Rowe[1,4]

*Corresponding author (kat35@cam.ac.uk, +44 1223 766 556, ORCID ID: 0000-0002-3178-6363)

[1] Department of Clinical Neurosciences, University of Cambridge, Cambridge, UK
[2] Department of Psychology, University of Cambridge, Cambridge, UK
[3] Department of Psychiatry, University of Cambridge, Cambridge, UK
[4] Medical Research Council Cognition and Brain Sciences Unit, Cambridge, UK

Summary: Accurate identification of brain function is necessary to understand the neurobiology of cognitive ageing, and thereby promote well-being across the lifespan. A common tool used to investigate neurocognitive ageing is functional magnetic resonance imaging (fMRI). However, although fMRI data are often interpreted in terms of neuronal activity, the blood-oxygen-level-dependent (BOLD) signal measured by fMRI includes contributions of both vascular and neuronal factors, which change differentially with age. While some studies investigate vascular ageing factors, the results of these studies are not well known within the field of neurocognitive ageing and therefore vascular confounds in neurocognitive fMRI studies are common. Despite over 10,000 BOLD-fMRI papers on ageing, fewer than 20 have applied techniques to correct for vascular effects. However, neurovascular ageing is not only a confound in fMRI, but an important feature in its own right, to be assessed alongside measures of neuronal ageing. We review current approaches to dissociate neuronal and vascular components of BOLD-fMRI of regional activity and functional connectivity. We highlight emerging evidence that vascular mechanisms in the brain do not simply control blood flow to support the metabolic needs of neurons, but form complex neurovascular interactions that influence neuronal function in health and disease.

*Keywords (3-6 words): neurovascular, cerebrovascular, cardiovascular, aging, fMRI, cognitive function*

# 1 Introduction

The worldwide population is rapidly ageing, creating a pressing need to understand the neurobiology of healthy cognitive ageing, over and above the problems associated with the rise of dementia in ageing societies (Beard et al., 2016). Understanding the neural mechanisms of healthy ageing will inform efforts to maintain cognitive function, which is critical for well-being across the lifespan (Sahakian, 2014). While neuroimaging has led to advances in knowledge about relationships between neural function and cognition, the effects of age on these interactions are poorly understood. This is due in part to outdated methodology, inadequate awareness and treatment of confounding variables, opaque reporting of results, lack of replication and a failure to consider the limitations of the signals of interest. In this paper, we review two complementary disciplines, neurocognitive ageing and neurovascular ageing, which have suffered from these limitations, and have proceeded somewhat independently. We argue for a better understanding of their relative contributions to functional magnetic resonance imaging (fMRI) signals, so as to formally integrate them in models of successful ageing, avoid common misinterpretations of fMRI and provide solutions to the limitations within each discipline alone.

The literature on neurocognitive ageing over the last 30 years has extensively relied on the blood-oxygen level-dependent (BOLD) signal detected for most fMRI. The fMRI signal reflects changes in deoxyhaemoglobin concentrations in response to neural activity (Figure 1). These concentrations change because increases in local synaptic activity and neuronal firing rates consume energy, which is sourced by transient local increase of cerebral blood flow (CBF) and cerebral blood volume (CBV). In simple terms, the dominant consequence is a temporary increase ("over-compensation") in oxygenated haemoglobin in the capillary and venous bed draining the activated region, reducing the concentration of deoxyhaemoglobin. Since deoxyhaemoglobin is paramagnetic, decreases in its concentration in turn increase the BOLD signal. The biophysical models of this *neurovascular coupling* include equations for dynamics of CBF, CBV and the cerebral metabolic rate of blood oxygen consumption ($CMRO_2$, for more details see Attwell and Iadecola, 2002; Hall et al., 2016; Kim and Ogawa, 2012). The resulting BOLD changes to a brief (<1s) period of neuronal activity can last up to 30s, with a characteristic temporal profile that is known as the *haemodynamic response function* (HRF) (Buxton et al., 2004; Friston et al., 1998a; Handwerker et al., 2012; Stephan et al., 2007). Many of the processes represented by parameters in these biophysical models are affected by ageing, owing for example to age-related changes in vascular health. Therefore, a failure to consider changes in vascular health can mean that differences in fMRI signals are erroneously attributed to neuronal differences (Hutchison et al., 2013a; Liu et al., 2013; Tsvetanov et al., 2015) and in turn their cognitive relevance misunderstood (Geerligs et al., 2017; Geerligs and Tsvetanov, 2016; Tsvetanov et al., 2016).

In this review, we first consider some of the main mediators of the transformation of neural activity into a haemodynamic response. We show how age-related alterations in the neuro-vascular interaction can influence the interpretation of changes in BOLD signal. This leads to changes in the measurements of regional activity and connectivity. We then turn to emerging evidence for the complex physiological changes with age, which give rise to slowing of cognitive function. These motivate the development of new models that characterise the joint contribution of vascular and neuronal influences to fMRI, in order to better understand the neurobiology of cognitive ageing. The continued interest in fMRI, above methods that are not affected by the vascular effects such as magneto- or electro-encephalography, rests on its safety, wide availability, high spatial resolution and full brain depth of imaging.

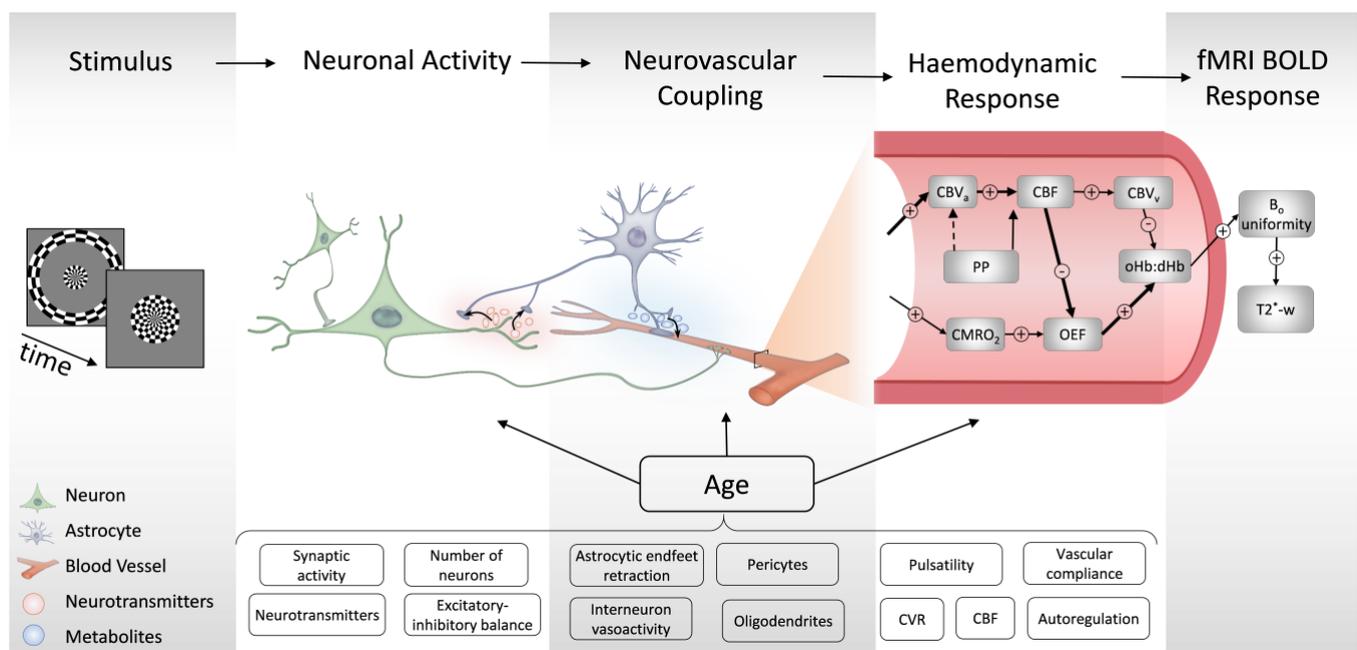

*Figure 1. A schematic illustration of the physiological basis of the BOLD response.* **Neuronal activity** elicited by a **stimulus** or background modulation gives rise to a complex **neurovascular coupling** signalling cascade. That triggers a **heamodynamic response** resulting in a blood-oxygen-level-depedent (**BOLD**) signal due to changes in the magnetic field inhomogeneity detected as a T2*-weighted signal by an MRI scanner. (Lower panel) Some of the suspected mediators of the differential age effects on the processes that give rise to the BOLD response. $CBV_a$ – arterial cerebral blood volume; $CBV_v$ – venous cerebral blood volume; CBF – cerebal blood flow; CVR – cerebral vascular reactivity; $CMRO_2$ – cerebral metabolic rate of blood oxygen consumption; oHb – oxygenated heamoglobin; dHb – deoxygenated heamoglobin; OEF – oxygen extraction fraction; $B_0$ – magnetic field;

# 2 Age-related changes in neuro-vascular influences

The study of neurovascular pathology has been relevant to understanding many medical, neurological and psychiatric disorders. Alterations in the neurovascular system during healthy ageing have also been studied at both the cellular and structural level. These changes typically remain undiagnosed and may have no directly apparent consequences for cognitive function, but they may compromise vasculature and the "neurovascular unit" that couple neuronal activity to vascular responses. This undermines the straightforward interpretation of BOLD as an index of neurometabolic activity in older populations, those on drugs that influence vascular function and many diseases that alter the neurovascular unit. The following section reviews the major neurovascular changes related to ageing, and considers the physiological consequences of structural changes for the BOLD fMRI signal.

## 2.1 Cellular and structural/morphological changes

### 2.1.1 *Vasculature, Blood vessels and the cerebrovascular tree*

Age leads to alterations in the cerebrovascular 'tree' at molecular, cellular and structural levels (Abdelkarim et al., 2019; Dai et al., 2012; Fulop et al., 2019; Paneni et al., 2017). Large elastic arteries dilate, stiffen and become atheromatous and tortuous, while the intima of the muscular arteries thickens (O'Rourke and Hashimoto, 2007). Vascular stiffening is also associated with alterations in smooth muscle cells (Lacolley et al., 2017), calcification and disruptions in the collagen-elastin balance (Kalaria and Hase, 2019; Kohn et al., 2015). Some arterial alterations are coupled with capillary rarefaction, in addition to molecular and morphological changes that perturb the brain-blood-barrier (BBB) (Abdelkarim et al., 2019; Kalaria and Hase, 2019; Li et al., 2018). Ageing is also associated with endothelial dysfunction, which contributes to dysregulation of vascular tone, astrocyte-dependent BBB permeability and nitric-oxide-dependent inflammation (Iadecola, 2017; Paneni et al., 2017). Damage to the endothelium aggravates vascular stiffening (Kalaria, 2010; Kalaria and Hase, 2019; Wang and Bennett, 2012) and compromises the vessels' ability to dilate and constrict in response to variations of blood pressure or vasoactive substances (Abdelkarim et al., 2019; Kohn et al., 2015). In addition, there is an age-related impairment in the mechanism underlying electrical propagation of retrograde hyperpolarization signal along the endothelial cells, thereby impairing the remote vasodilation of upstream pial arterioles and increased perfusion in the capillary bed (Iadecola, 2017).

Pericytes are a group of microvascular mural cells embedded in the basement of blood microvessels that regulate blood flow both physiologically and pathologically (Hall et al., 2014) in addition to fine tuning vascular tone and BBB permeability with their contractile properties (Attwell et al., 2016; Sweeney et al., 2016). Age-related changes in pericytes (Abdelkarim et al., 2019; Kalaria and Hase, 2019) together with other mural cell alterations are likely to lead to changes in the vascular basis of the BOLD signal.

In short, disruption of a myriad of cerebrovascular factors acting on different levels of the vascular tree contributes synergistically to changes in neurovascular signalling, perfusion and reactivity.

*2.1.2   Neuronal and non-neuronal cells*

Neurons can directly control cerebral blood flow (Drew, 2019). In particular, interneurons produce vasodilators (Lee et al., 2019; Uhlirova et al., 2016) and vasoconstrictors (Cauli and Hamel, 2010), such as nitric oxide, prostanoids, endothelin etc (for more informatino on vasoactive agents see Hosford and Gourine, 2019; Iadecola, 2017). Stimulation that selectively targets interneurons causes a relatively small increase in oxygen consumption but a relatively large increase in CBF. In contrast, stimulation of excitatory neurons causes relatively large increase in oxygen consumption but relatively small increase in CBF (Echagarruga et al., 2019; Vazquez et al., 2018). These results suggest that, while the primary driver of the BOLD response (i.e. CBF) is interneuron activation, additional $CMRO_2$-mediated changes in BOLD signal reflect excitatory neuron-modulated oxygen consumption (Drew, 2019). The interpretation of these findings in the context of ageing is important, given the dissociating effects of age on excitatory *versus* inhibitory signalling and synapses (Legon et al., 2016; Luebke et al., 2004; McQuail et al., 2015; Rozycka and Liguz-Lecznar, 2017).

While the role of glial cells in the neurovascular unit is less well understood than neuronal and vascular components, increasing evidence implicates glial elements as mediators between neurons and blood vessels (Sweeney et al., 2019). Astrocytes are a diverse population of glial cells whose functions include neurovascular signaling (Khakh and Deneen, 2019), linking neurons to their blood supply (Zonta et al., 2003) and regulating the BBB (Petzold and Murthy, 2011). Activated astrocytes release vasoactive agents via multiple signaling pathways at different levels of the vascular tree (Cauli and Hamel, 2018) independently from other endothelial pathways (Chen et al., 2014) including caveolea-mediated vasodilation in arterial endothelium (Chow et al., 2020). Retraction of the astrocytic endfeet as part of the clasmatodendrotic response (Chen et al., 2016), together with changes in the immune response and calcium signaling, impairs byproduct clearance as the BBB efficiency breaks down (Abdelkarim et al., 2019; Kress et al., 2014). Age-related changes in glia (Abdelkarim et al., 2019; Kalaria and Hase, 2019) together with mural and endothelial cell alterations could be stronger than that of neurons (Soreq et al., 2017) and are likely to lead to changes in the vascular basis of the BOLD signal. Future work needs to consider how glial elements could be measured, integrated with in-vivo neuroimaging and accounted for in physiologic ageing models.

While both glia and neurons play a role in the vascular basis of the BOLD signal, their relative contribution to baseline (endogeneous) BOLD signal *versus* evoked (exogeneous, e.g, task-based) BOLD remains unclear. This is important because the baseline of blood flow, which decreases with ageing, can affect the sign and the magnitude of the evoked BOLD signal, without changes in underlying neural activity (Brown et al., 2003; Cohen et al., 2002; Stefanovic et al., 2006; Takata et al., 2018). Age may differentiate the separate factors that regulate artery tone (Drew, 2019) *versus* evoked responses (Echagarruga et al., 2019). Taken together, it appears that subtypes of glia, mural cells, endothelium and inter-neurons control the BOLD signal, independent of the activity of the neighbouring excitatory neurons, and likely through multiple signalling mechanisms that contribute synergistically to vasodilation. Multiple NVC pathways acting on different levels of the vascular tree crucially depend on well-orchestrated interplay between different cell types of the neuro-glio-vascular unit, which may provide multiple safety mechanisms (Cauli and Hamel, 2018). Ultimately, an integrated understanding of age effects on all components of the neuro-glio-vascular unit is required for a better understanding of the physiological basis of neurocognitive ageing, especially where inferences are drawn from fMRI.

## 2.2 Physiological changes

It is the effects of age on cerebrovascular function that render interpretation of age differences in the BOLD signal so challenging. Cerebrovascular function can be assessed by measuring: 1) resting cerebral blood flow (CBF), 2) CBF responses to changes in arterial CO2, referred to as cerebrovascular reactivity (CVR), 3) CBF responses to changes in blood pressure, referred to as cerebral autoregulation, and 4) CBF responses to changes in neural activation, referred to as neurovascular coupling (NVC). Cerebrovascular alterations also include brain pulsatility and the cerebral metabolic rate of oxygen extraction. Below we review of these changes based on the common range of physiological recordings (see also Nagata et al., 2016; Iadecola, 2017; Toth et al., 2017; Abdelkarim et al., 2019).

### 2.2.1 Resting cerebral blood flow

Decrease in global baseline CBF with age has been reported in early studies using transcranial Doppler ultrasonography (Flück et al., 2014), radiotracer techniques (Leenders et al., 1990; Nagata et al., 2016; Reich and Rusinek, 1989) and phase contrast imaging (Ambarki et al., 2015). These changes are widespread across the cerebral cortex and the basal forebrain. The physiology underlying the CBF decrease in the aged brain is still debated (Tarumi and Zhang, 2018). The main candidates include primary causes of impaired vasoactivity and cardiovascular regulation of CBF during ageing, rather than the reduction in cardiac output (Xing et al., 2017). CBF decline may also reflect the secondary effects of brain atrophy and reduction in neural activity as a shift towards lower metabolic demands, rather than primary changes in vasculature. The finding that changes in CBF can affect the sign and magnitude of the evoked BOLD signal without affecting underlying neural activity (Brown et al., 2003; Cohen et al., 2002; Stefanovic et al., 2006) is in line with the deoxyhaemoglobin-dilution model (Davis et al., 1998; Hoge et al., 1999a, 1999b). Therefore, the decline in the baseline CBF with ageing has implications for the interpretation of fMRI studies of ageing.

### 2.2.2 *Cerebrovascular reactivity*

Cerebrovascular reactivity (CVR) is informative about vascular health. CVR is distinct from resting CBF, as it measures the ability of cerebral arteries and arterioles to dynamically regulate blood supply through dilation or constriction. In particular, CVR reflects the CBF responses to changes in arterial $CO_2$, whereby elevated partial pressure of arterial $CO_2$ (hypercapnia) causes dilation of vascular smooth muscle, leading to regional increases in CBF, while reduced $CO_2$ partial pressure (hypocapnia) causes vasoconstriction leading to regional decreases in CBF. Vascular sensitivity to $CO_2$ is very marked in the cerebrovasculature (Ainslie et al., 2005) and is thought to depend on intra- and extracellular pH changes that modulate vascular smooth muscle tone (Jensen et al., 1988; Lambertsen et al., 1961; Lassen, 1968). Therefore, CVR is considered to be a more direct measure of vascular endothelium and smooth muscle function compared to baseline CBF. $CO_2$ quantification in cerebrovasculature has used transcranial Doppler ultrasound (Ainslie and Duffin, 2009), radiotracer techniques (Ito et al., 2001) and contrast imaging (Chen et al., 2006). There is general agreement across multiple imaging techniques that changes in CBF relative to changes in $CO_2$ partial pressure are similar between brain regions under hypercapnia, but not under hypocapnia (Willie et al., 2014). Experimental modulation in $CO_2$ partial pressure has been used to validate non-invasive perfusion techniques (Rostrup et al., 1996, 1994), as well as biophysical (Keyeux et al., 1995) and biochemical (Wagerle and Mishra, 1988) aspects of cerebral vasodilation.

Global decline in CVR with age has been reported using transcranial Doppler ultrasound (Flück et al., 2014), radio tracer techniques (Reich and Rusinek, 1989; Tsuda and Hartmann, 1989; Yamaguchi et al., 1979) and phase contrast imaging (Geurts et al., 2018). Age-related differences in the response of regional CBF to $CO_2$ inhalation have been reported using PET (Ito et al., 2002). Reduction in hypercapnia-induced vasodilation in the cerebellum and insular cortex, as well as hypocapnia-induced vasoconstriction in the frontal cortex, has been observed in older adults, suggesting less effective vascular response in cerebral perforating arteries (Ito et al., 2002). Likely causes for CVR changes are arterial stiffening (Fang, 1976) and compromised endothelial function in blood vessels (Brandes et al., 2005), which lead to a decreased vascular response to match metabolic demands. In addition, white matter hyperintensities, a common MRI finding in ageing, are associated with reduced baseline CBF and reduced response to hypercapnia (Hatazawa et al., 1997; Kuwabara et al., 1996). Compromised

CVR will lead to a reduced dynamic range of the BOLD signal, having direct implications for task-based fMRI studies of ageing: even with the same levels of neural activation across age groups, lower CVR in the older group would lead to smaller amounts of vasodilation and therefore reduced evoked CBF, reduced decrease of deoxyhaemoglobin concentration and reduced BOLD signal. Without controlling for CVR differences, this would lead to an under-representation of neural responses in older individuals.

### 2.2.3 Pulsatility

Cyclic cardiac contractions that pump blood through the arterial system generate a pulsatile blood flow and concomitant pulsatile pressure experienced by vascular wall tissue. This pulsatile phenomenon is absorbed before it reaches pressure-sensitive cerebral capillaries, and maintenance of steady flow and pressure ensures exchange of nutrients and clearance by-products. The first line of defence to minimise the effect of flow and pressure pulsatility in the microcirculation is achieved by the highly elastic aorta and muscular arteries, e.g. the aorta-carotid interface. The distensibility mismatch in these vessels dampens the pulsatile energy projected distally, known as the Windkessel effect (Mandeville et al., 1999; Wagshul et al., 2011). Arterial stiffening caused by imbalance in elastin-collagen in the load-bearing intima of the aorta and central elastic arteries alters arterial distensibility, translating into increased pulse wave velocity (Vlachopoulos et al., 2011). The change of pulse wave velocity with age alters the wave reflection properties at the aorta-carotid interface, resulting in less effective cushioning of pulsations in the arterial system, i.e. diminished Windkessel effect, and greater transmission of pulsatile energy into the cerebral microcirculation (Nagata et al., 2016). Increase of pulsatility in the proximal part of cerebrovasculature is further exacerbated with age increase in pulse pressure (increased difference between systolic and diastolic pressure). Transcranial Doppler ultrasound of major arteries entering the brain, together with phase contrast MRI of the whole brain, both point to an age-related increase in cerebral pulsatility (Mitchell et al., 2011; Tarumi et al., 2014). These changes can potentially contribute to microvascular ischaemia and tissue damage that is seen in some MRI-derived measures (Mitchell et al., 2011; Mok et al., 2012; Webb et al., 2012). These microvascular changes have in the past been considered as a benign feature of ageing, but may actually be a significant contributor to changes in neurocognitive function (Wåhlin and Nyberg, 2019). With regards to BOLD imaging, pulsatile blood flow not only leads to fluctuation in signal intensity in arteries, arterioles and other large vessels (Dagli et al., 1999), but also an age-related increase in pulsatility deeper in microvasculature. This could have dramatic effects on the BOLD signal in the proximity of neuronal tissue, which has only recently been recognised as a potential confound of BOLD studies (Makedonov et al., 2016, 2013; Theyers et al., 2018; Tsvetanov et al., 2015; Viessmann et al., 2019, 2017).

*2.2.4   Cerebral autoregulation*

The second line of defence for minimising pressure fluctuations in brain microvasculature is cerebral autoregulation, referred to here as autoregulation, via the vessels' ability to dilate or constrict in response to systemic perfusion pressure changes (Willie et al., 2014).This is complementary to CVR (Battisti-Charbonney et al., 2011; Jeong et al., 2016). In particular, autoregulation constitutes the ability of the cerebral vasculature to maintain steady flow and pressure in the capillary bed during transient changes in arterial pressure or intracranial pressure. The myogenic response, which is intrinsic to the vascular smooth cells and a key mechanism to autoregulation, is impaired with ageing, especially under conditions of hypertension and increased pressure pulsatility (Toth et al., 2017). Furthermore, impaired autoregulation precedes vascular damage in white matter (Joutel et al., 2010) and relates to white matter hyperintensities (Purkayastha et al., 2014) in animal and human studies, respectively.

The CVR and autoregulation adjustment of vascular resistance to varying arterial $CO_2$ and pressure, respectively, is primarily modulated in large arteries and pial arterioles (Willie et al., 2014). This is suggesting that BOLD-related measures targeting CVR and autoregulation may be sensitive to ageing effects in the proximal part of the vasculature, and less sensitive to independent changes in the distal part of the cerebral circulation and physiological factors therein. In other words, CVR and autoregulation may be less sensitive to mechanisms underlying retrograde intramural propagation of vascular signals, causing remote vasodilation of upstream pial arterioles, i.e. impairing mechanisms of communication within the neurovascular unit (Chen et al., 2014).

## 2.3   Drug effects

Older people are more likely to be taking medications, including medications for age-related chronic disorders such as high blood pressure, diabetes, blood clotting, arthritis or neurodegeneration. Ageing studies often do not explicitly address these potential confounds in the interpretation of their results, which potentially modifies their conclusions, especially since the drugs are likely to affect the cascade of signalling and vascular events that form the basis of the BOLD signal (Iannetti and Wise, 2007). Approaches aimed at dissociating vascular from neuronal signals should seek to identify, characterise and control for the confounding effects of drugs in ageing studies.

# 3 Dissociating neuro-vascular influences in BOLD fMRI signal

In order to interpret fMRI data, one needs to understand the contributions of neuronal and vascular components of signal variance. Some studies of neurocognitive ageing attempt to bypass the impact of vascular influences through their inclusion criteria, e.g. excluding individuals with a history of hypertension, cardiovascular or neurological conditions. These are, however, categorical criteria, and insensitive to continuous variation in the population, and unable to resolve the effects of undiagnosed/presymptomatic conditions; not to mention producing results that are potentially biased, in not generalising to the typical ageing person.

There are several other approaches to separate, or "unconfound", neural from vascular factors. These are summarised as three broad strategies (Figure 2). The first is based on detecting vascular signals in BOLD fMRI data by using independent measurement of vascular signals, termed here *vascular unconfounding*. The second relies on identifying neuronal signals in BOLD fMRI by using independent measurement of neuronal signals, termed *neuronal integration*. The third uses formal *modelling approaches* to the fMRI signal. Below we review the methods and exemplar applications within each strategy, and illustrate their strengths and weaknesses for studying the effects of ageing.

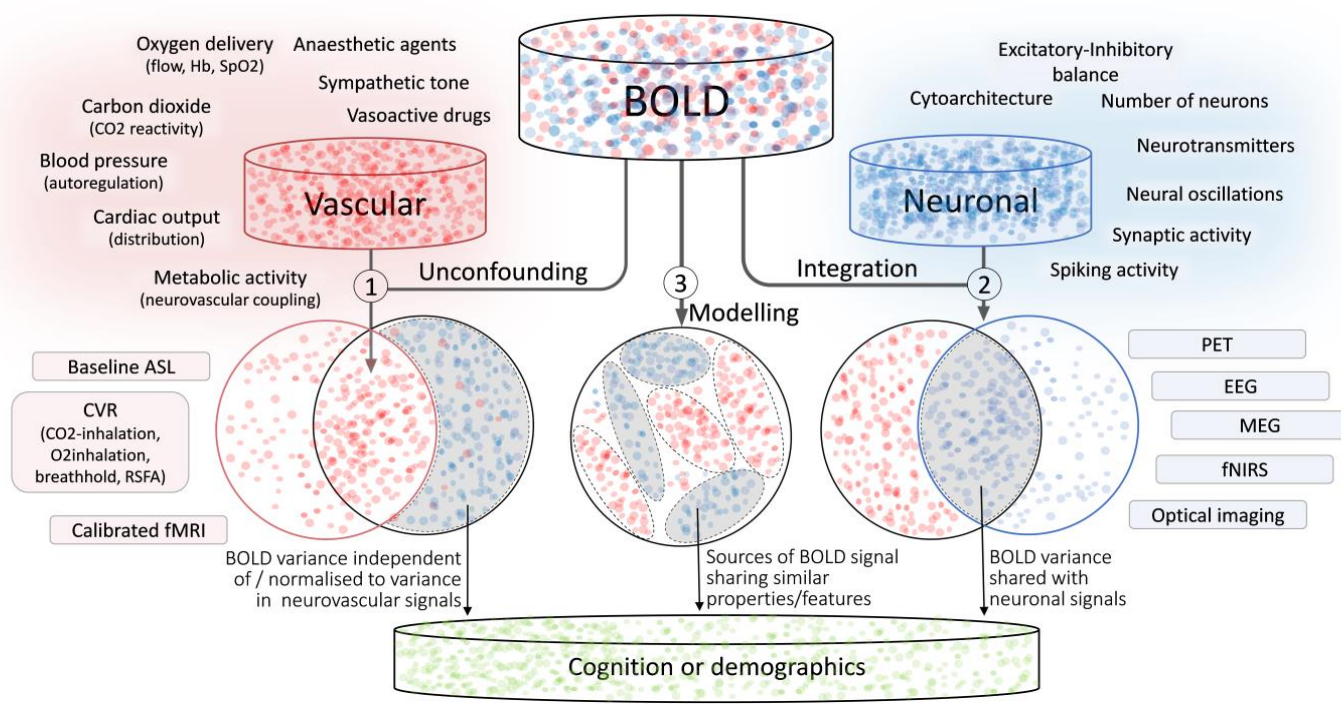

*Figure 2. Schematic illustration of dissociating neurovascular influences in fMRI-BOLD using (1) vascular unconfounding, (2) neuronal integration and (3) BOLD modelling.*

## 3.1 Vascular detection using fMRI

The first class of approaches focuses on estimating vascular contributions to the BOLD signal using an independent MRI-based measurement that aims to capture individual variability in one or more of the physiological factors discussed in the previous section. An implicit assumption of these methods is that they explain variability in vascular signals (see figure), but not variability related to neural activity, such that they can be used to "adjust" the BOLD signal without removing age-related neuronal changes. These approaches further fall into calibration and normalisation methods. We focus here on task-evoked BOLD responses in voxel-wise imaging, but the principles can be applied to other forms of analysis.

### 3.1.1 Normalization using baseline cerebral blood flow

Regional baseline CBF has for many years been measured with positron emission tomography (PET) (Mintun et al., 1984) or MRI using tracer kinetic procedures (Østergaard, 2005). However, safety concerns associated with tracers and the complexity of procedures have limited their application in BOLD studies of ageing. The predominant fMRI method for estimating resting CBF is based on endogenous contrast generated through perfusion of blood water into brain tissue (Belliveau et al., 1991). The signal intensity is generated by applying a magnetic label to proton spins of the inflowing arterial blood water, termed arterial-spin labelling (ASL) (Detre et al., 2009). In analogy to PET perfusion imaging, the ASL "tracer" is the endogenous arterial blood water, where the magnetic label decays with T1 instead of radioactive decay. Deep understanding of the physiological basis of ASL and validation against radiotracer approaches (Chen et al., 2008; Detre et al., 1990; Eleff et al., 1988; Heijtel et al., 2014; Hua et al., 2019; Pekar et al., 1991) has established ASL as a robust and non-invasive technique to provide quantitative estimate of baseline CBF. An overview of ASL-variations and an agreement on its application has been discussed previously (Alsop et al., 2015).

Resting state ASL studies of ageing support the presence of age-related atrophy-independent decreases in resting CBF throughout the cortex (Leoni et al., 2017; Tsvetanov et al., 2019b; Zhang et al., 2017). Some studies also suggest a non-linear effect across the lifespan (Biagi et al., 2007). Interestingly, increases in regional CBF, in lateral and medial temporal lobe for example, have been observed with increasing age (Preibisch et al., 2011; Salat et al., 2009; Tsvetanov et al., 2019b), which may reflect macro-vascular artifacts (Detre et al., 2012; Mutsaerts et al., 2017) due to prolonged arterial transit time with ageing (Dai et al., 2017).

ASL studies support the proposal that age-related decline in baseline CBF reflects both cardiovascular and neurovascular impairment (Zhang et al., 2017). For example, age-related reduction in baseline CBF occurs in cortical regions typically associated with high vascular risk and genetic factors (Bangen et al., 2014; Ighodaro et al., 2017), and precedes brain atrophy (Hirao et al., 2006; Matsuda et al., 2002; Ruitenberg et al., 2005). However, recall that baseline brain perfusion is highly dependent on other physiological factors, and the difference in CBF may also reflect age bias in these factors, rather than baseline changes in CBF (Grade et al., 2015). For example, baseline ASL may reflect spontaneous $CO_2$ fluctuations, medication use, time of day, or levels of wakefulness (Clement et al., 2018). Some influences are global and related to vascular tonus, while other local variations are the result of psychotropic effects on the brain. As an example, physical exercise, drinking coffee or smoking just before the perfusion measurement have substantial influence on both global and local quantification (Addicott et al., 2009; Domino et al., 2004; MacIntosh et al., 2014; Merola et al., 2017). While this may be a drawback for absolute CBF quantification, it is an advantage for the use of ASL as a normalisation technique, i.e, to control for multiple physiological factors in BOLD studies.

ASL has been broadly used to estimate baseline CBF, and has been used to try to rule out drug effects on vascular contributions to BOLD effects of disease or drug (Macoveanu et al., 2013; Rae et al., 2016). However, it is rarely used for a formal normalizing approach in evoked BOLD studies of age. This could be due to the low signal-to-noise ratio of ASL, low spatial resolution and additional time needed to acquire baseline CBF, and a preference to integrate it within a BOLD fMRI acquisition (see below). In one study, regional age-related differences in BOLD activation were shown to be mediated by baseline ASL-CBF, suggesting a substantial vascular contribution with regional specificity to the observed BOLD age-differences (Zebrowitz et al., 2016). In summary, the improvement in quality and application of baseline ASL-CBF measurements in the recent years offers advantages over some of the other following approaches to control for age-related differences in physiological influences of BOLD signal.

*3.1.2 Normalization using cerebrovascular reactivity*

This approach differs from the baseline perfusion approach in that it relies on experimentally-perturbed physiological states during the MRI scan. This physiological response, defined as cerebrovascular reactivity above, leads to changes in BOLD signal that are dominated by vascular factors (reflecting transient variations in physiological factors) in the absence of apparent changes in neuronal activity. In particular, estimation of cerebrovascular reactivity exploits the molecular mechanisms of $CO_2$-induced vasodilation (discussed above), which can be used to model variability in physiological signals of evoked BOLD data. Fluctuations in arterial-blood $CO_2$ can take three forms of hypercapnia: $CO_2$ administration, voluntary breathhold or naturally occurring fluctuations linked to respiration during a resting state fMRI acquisition (discussed below and in Liu et al., 2019). As the CVR manipulations work under the assumption of no changes in the underlying neuronal activity and oxygen extraction ($CMRO_2$) (Davis et al., 1998), the CVR manipulation takes the form:

$$\Delta BOLD_{CVR} = M(1 - f_{CVR}^{\alpha-\beta})$$

where $f$ = $CBF/CBF_0$ represents CBF signal normalised by its respective baseline value. The subscript *CVR* denotes the hypercapnia condition and the parameter *M* defines the maximum possible BOLD signal change for a brain region. The superscript parameters are determined empirically, but are well approximated as α ≈ 0.4 and β ≈ 1.5 (Buxton et al., 2004). Dividing the task-based BOLD response by the hypercapnia response yields a normalized BOLD response of the form:

$$\Delta BOLD_N = \frac{(1 - f_F^{\alpha-\beta} m_F^{\beta})}{(1 - f_{CVR}^{\alpha-\beta})}$$

where $m$ = $CMRO_2/CMRO_{2,0}$ represents $CMRO_2$ signal normalised by its respective baseline value. The subscript *N* denotes the normalised BOLD response. Note that the *M* term cancels out in the normalized response, which precludes estimation of modulatory factors of *M*, such as magnetic field strength and baseline blood volume and oxygenation (Bandettini and Wong, 1997). This normalisation procedure entails the division of a functional contrast map by a cerebrovascular reactivity (CVR) map, i.e. normalisation at each voxel. However, including the hypercapnia response as a covariate in a voxel-level, general linear model (GLM), together with the functional BOLD response (e.g. when predicting behavioural or demographic measures) might provide a better approach (Liau and Liu, 2009). It is worth noting that these operations assume a linear relationship between CBF and BOLD signal that holds across varying $CO_2$-levels in arterial blood. However, this may not always be true (Halani et al., 2015), given claims of a non-linear BOLD-CBF relationship (Hoge et al., 1999c) and a non-linear response of vasculature to large $CO_2$ and arterial pressure fluctuations (Battisti-Charbonney et al., 2011; Mitsis et al., 2004). This is further complicated by the "vascular steal" phenomenon of flow diversion from regions of low to high cerebrovascular reactivity (Sobczyk et al., 2014) and interactions between multiple physiological factors that increase with age (Kalaria and Hase, 2019). While this warrants future research on modelling the non-linear nature of the effects, the current approaches may lead to underestimation, rather than overestimation, and therefore still offer a partial solution to minimise vascular influences in evoked BOLD signal.

### 3.1.2.1 *CO2-induced hypercapnia*

An individual's hypercapnia response can be modulated by inhaling a special gas mixture inside the MRI scanner. Bandettini and Wong (1997) were the first to demonstrate the utility of this technique for BOLD fMRI studies. For a technical review and practicalities of this approach using various types of apparatus, see Liu et al. (2012, 2019) and Germuska and Wise (2019). Regardless of the gas-delivery apparatus, accurate assessment of CVR relies on tracking the maximal concentration of CO2 in the exhaled air - so-called "end-tidal CO2" (Et-CO2) - during breathing cycles with varying CO2 concentration in the inhaled gas (see in video format, Lu et al., 2014). The variation of Et-CO2 is tightly linked to changes in alveolar pressure of CO2 and fluctuations in arterial vasodilation, indicating the extent to which the vascular system is challenged. The analysis of CO2-induced hypercapnia data is conceptually similar to the task-evoked fMRI, where the temporally-aligned Et-CO2 timecourse is included as the main regressor in a GLM to produce cerebrovascular response ($CVR_{CO2}$) brain map. Early studies of $CO_2$-induced cerebrovascular reactivity demonstrate a close regional overlap between voxels showing BOLD $CO_2$ responses (BOLD-$CVR_{CO2}$ map) estimated from fMRI and voxels showing a cerebrovascular response (CBF-$CVR_{CO2}$ map) estimated from PET (Rostrup et al., 2000) and ASL (Halani et al., 2015; Hare et al., 2013; Zhou et al., 2015). Although this overlap has high reproducibility (Kassner et al., 2010) across various field strengths and MRI sequences (Cohen et al., 2004), BOLD-$CVR_{CO2}$ is more sensitive to basal CO2 fluctuations than CBF-$CVR_{CO2}$ (Halani et al., 2015).

BOLD-$CVR_{CO2}$ in grey matter declines with age (De Vis et al., 2015; Liu et al., 2013; Lu et al., 2011; Thomas et al., 2014; Zhou et al., 2015). The age effects on BOLD-$CVR_{CO2}$ are more prominent than those on baseline ASL-CBF (De Vis et al., 2015; Lu et al., 2011) and exhibit distinct regional patterns (Leoni et al., 2017), supporting the notion of age having independent effects on CVR and baseline CBF (Tsvetanov et al., 2019b). Interestingly, age-related BOLD-$CVR_{CO2}$ increases are found in white matter (Thomas et al., 2014), which may reflect changes in the mechanical properties of the white matter: White matter in older adults becomes less densely packed due to demyelination and axon loss, making it easier for blood to penetrate and vessels to dilate.

Combining BOLD-$CVR_{CO2}$ with evoked BOLD studies of ageing allows correction for regionally-specific effects (Gauthier et al., 2013; Liu et al., 2013), which could lead to improved associations between BOLD estimates and outcomes of interest (Song et al., 2016). For example, age-related decreases in evoked BOLD responses in V1 and medial temporal lobe were abolished after correction, while age-related increases in bilateral frontal gyrus remained after correction. This suggests that many age-related differences found in fMRI studies reflect changes in vasodilation rather than in neuronal activity.

Unfortunately, such corrective methods have not been widely used, in part due to impracticalities of large-scale studies, and tolerance by older adults and clinical populations (Liu et al., 2012a; Spano et al., 2013). Furthermore, a gas-induced hypercapnic challenge may not be neuronally neutral (Driver et al., 2016; Golestani et al., 2016; Marshall et al., 2015; Xu et al., 2011), e.g, given participants' awareness of the aversive challenge, and this effect on neural activity may differ with age (Hall et al., 2011). In this case, correction by $CVR_{CO2}$ might obscure true task-related neural differences with age. Nonetheless, recent developments in the gas challenge procedure allow for estimation of multiple physiological parameters, including venous oxygenation and resting state functional connectivity (see below), which may improve the accuracy of corrections for vascular signals in BOLD fMRI studies.

### 3.1.2.2 Breath-hold-induced hypercapnia

An alternative way to modulate arterial CO2 in the absence of gas-delivery apparatus involves breathing challenges, where participants endogenously increase arterial $CO_2$ by voluntarily holding their breath (Kastrup et al., 1999), which we term $CVR_{BH}$ to indicate breath-holding (Urback et al., 2017). BOLD-based $CVR_{BH}$ demonstrates high correspondence with ASL-based $CVR_{BH}$ (Bulte et al., 2009; Fukunaga et al., 2008), BOLD-$CVR_{CO2}$ (Handwerker et al., 2007; Kastrup et al., 2001; Murphy et al., 2011; cf. Tancredi and Hoge, 2013) and has excellent repeatability (Bright and Murphy, 2013; Lipp et al., 2015). Improved $CVR_{BH}$ estimation may be achieved using variations in the breath-hold procedure (Bright et al., 2009; Magon et al., 2009; Scouten and Schwarzbauer, 2008; Thomason and Glover, 2008) and in analysis of the data (Bright and Murphy, 2013; Pinto et al., 2016; van Niftrik et al., 2016).

Beyond its use to minimise inter-individual variability of physiological influences in BOLD studies of young adults (Chang et al., 2008; Di et al., 2012; Handwerker et al., 2007; Kannurpatti et al., 2011b; Liau and Liu, 2009; Murphy et al., 2011; Thomason et al., 2007), breath-holding has been used more commonly in ageing studies than other normalisation approaches (Di Luft et al., 2016; Gonzales et al., 2014; Handwerker et al., 2007; Kannurpatti et al., 2011a, 2010; Mayhew et al., 2010a; Mayhew and Kourtzi, 2013; Riecker et al., 2003). Riecker and colleagues showed that the age-differences in BOLD response of sensorimotor regions during finger tapping were accompanied by differences in BOLD-$CVR_{BH}$ (Riecker et al., 2003), which was one of the first indications that evoked fMRI studies of ageing require careful interpretation of observed BOLD differences. Later studies extended these findings to other primary sensory regions, and corroborated the idea that age-related decline in evoked BOLD response to sensorimotor stimuli can be accounted for by age differences in $CVR_{BH}$ (Handwerker et al., 2007; Kannurpatti et al., 2010). Interestingly, age differences in BOLD signal in "higher-order" cortical regions during cognitive tasks often remained after controlling for $CVR_{BH}$, suggesting the relationship between BOLD signal, neural activity, vascular signal and age varies across brain regions (Handwerker et al., 2007; Kannurpatti et al., 2010). More recent studies confirm that consideration of $CVR_{BH}$ not only changes the pattern of regional age-differences in evoked BOLD response (Mayhew et al., 2010; Di Luft et al., 2016), but also improves the strength of the relationship between BOLD responses and performance on the task (Raut et al., 2016).

However, while breath-holding may be more tolerable and has been employed in more ageing studies than gas-induced CVR, the compliance to breath-holding procedure, lung capacity, inspiration and expiration ability of participants may decrease with their age (Jahanian et al., 2017). Such biases affect data quality and reliability measures (Magon et al., 2009; Tancredi and Hoge, 2013), highlighting the advantage of other less invasive ("task-free") estimates of vascular components of BOLD time series.

### 3.1.2.3 Resting State Fluctuation Amplitudes

One such "task-free" estimate of the vascular component of the BOLD signal is the intrinsic variability of the BOLD signal across time (after bandpass filtering to remove slow drifts in MR signal and high frequency motion artifacts). This is known as resting state fluctuation amplitude (RSFA). Early studies demonstrate that RSFA reflects naturally occurring fluctuations in arterial $CO_2$ induced by variations in cardiac rhythm and in respiratory rate and depth (Golestani et al., 2015; Wise et al., 2004). RSFA approximates the BOLD response to hypercapnic challenge and was proposed as a safe, scalable and robust cerebrovascular reactivity mapping technique (Kannurpatti and Biswal, 2008; Tsvetanov et al., 2015). As with other methods discussed above, the use of RSFA as a correction method for BOLD requires the assumption that age differences in RSFA reflect only vascular factors, rather than age-related differences in neural function. Although RSFA demonstrates high correspondence across brain regions and individuals when compared to baseline ASL-CBF (Garrett et al., 2017; Tsvetanov et al., 2019b), BOLD-$CVR_{BH}$ and BOLD-$CVR_{CO2}$ (Jahanian et al., 2017; Kannurpatti and Biswal, 2008; P. Liu et al., 2012b), and in groups with compromised CVR (De Vis et al., 2018; Jahanian et al., 2014), the effects of age on RSFA cannot be fully explained by these factors (Garrett et al., 2017; Tsvetanov et al., 2019b). Therefore, without understanding the unexplained effects of age on RSFA in terms of neuronal versus vascular influences it would be dangerous to use RSFA as a normalisation technique.

Recent evidence has improved our understanding of the origins of RSFA. For example, pulsatile effects could influence the BOLD signal in the proximity of large brain vessels and CSF. In particular, the age-related increase in pulsatility deeper in microvasculature (see *pulsatility* Section 2.2.3) is may likely to contribute to the RSFA signal, as recently recognised (Makedonov et al., 2016, 2013; Theyers et al., 2018; Tsvetanov et al., 2015; Viessmann et al., 2019, 2017). This could explain why Tsvetanov and colleagues found that age differences in RSFA are either fully or partly mediated by heart rate variability. In contrast, these authors further found no evidence that neural variability (as measured by magnetoencephalography) mediated the age effects of RSFA (Tsvetanov et al., 2015); findings further supported by EEG-based neural estimates (Kumral et al., 2019). However, while age-related differences in RSFA may not reflect neuronal signals, the use of either somatic vascular measures or cerebrovascular measures explained only part of the age-related differences in RSFA. This leaves open the possibility that age-related differences in RSFA reflect joint contributions from cardiovascular and neurovascular factors, as in the case of BOLD signal fluctuations (Chang et al., 2013, 2009).

To resolve this ambiguity, we followed up our original study by considering the simultaneous assessment of the independent and shared effects of cardiovascular, neurovascular and neuronal effects on age-related differences in RSFA (Tsvetanov et al., 2019b). After controlling for either cardiovascular and neurovascular estimates alone, the residual variance in RSFA across individuals remained significantly associated with age, replicating the above findings. However, when controlling for both cardiovascular and neurovascular estimates, the residual variance in RSFA was no longer associated with age. This suggests that cardiovascular and neurovascular signals are together sufficient predictors of age-related differences in RSFA. In summary, while originally proposed to control for CVR (Kannurpatti and Biswal, 2008), RSFA captures multiple vascular signals that are independently affected by age, and appears to be a valid method to correct for vascular factors in the BOLD signal, in order to better characterise effects of age on neural, and ultimately cognitive, function.

When RSFA is used to correct evoked BOLD data, the amplitude and spatial pattern of the normalised response are similar to that when using $CVR_{BH}$ and $CVR_{CO2}$ (Kannurpatti and Biswal, 2008). Controlling for RSFA has been shown to minimise non-neural BOLD variability across individuals (Kannurpatti et al., 2011b), in populations with impaired cardiovascular health (Nair et al., 2017; Ravi et al., 2019), and improve estimation of evoked BOLD signals related to distinct neuronal mechanisms (Mayhew et al., 2016). In studies of ageing, controlling for RSFA in evoked BOLD signal accounts for age-related differences in BOLD response in some sensory regions, comparable to findings from alternative normalisation approaches (Kannurpatti et al., 2014, 2011a, 2010; Tsvetanov et al., 2015). Importantly, not all age differences disappear after controlling for RSFA – for example, the ipsilateral motor cortex overactivation in older adults remains, consistent with results from other approaches used to study ageing effects on the motor system (Boyke et al., 2008; Rowe et al., 2006; Scholz et al., 2009; Tigges et al., 1990).

Variations in the estimation of RSFA exist, which may be more sensitive to CVR relative to cardiovascular signals (Liu et al., 2017). In addition, other means of RSFA-like estimates have been proposed to derive from non-resting cognitive states (Kazan et al., 2016) or fixation-/resting-periods succeeding task periods (Garrett et al., 2017). Given that short periods of cognitive engagement have been shown to modulate the BOLD signal in a subsequent resting state scan (Sami et al., 2014; Sami and Miall, 2013), future studies are required to generalise RSFA findings to RSFA-like estimates derived from other types of fMRI acquisitions.

- *Other CVR-induced variations*

The measures discussed above could be complemented with other physiological measures, such as total baseline venous oxygenation from phase contrast MRI (Lu et al., 2010), which may provide superior SNR compared to ASL, or cerebral blood volume (Hua et al., 2019) with multiple physiologic parameters (Bright et al., 2019). Their usefulness for estimating age-related differences in the vascular component of the BOLD signal remains to be demonstrated.

### 3.1.3 Calibration using concurrent fMRI

The use of so-called "calibrated fMRI" has been possible for many years and can in theory control for differences in both baseline physiology and haemodynamic coupling across individuals and hence ages. The technique involves measuring blood flow, blood volume and venous oxygenation via normocapnic and hypercapnic/hyperoxic gas challenge during a concurrent measurement of BOLD and CBF (Blockley et al., 2015; Germuska and Wise, 2019; Pike, 2012). As such, calibrated fMRI provides a measure of relative changes in $CMRO_2$, which can be integrated with physiological models (Gauthier and Fan, 2019; Gauthier and Hoge, 2012), to estimate quantitatively the absolute rate of cerebral metabolic oxygen consumption ($CMRO_2$), i.e. oxidative metabolism, from the data. However, it has not seen widespread adoption, mainly because it is difficult to implement.

In theory, measurements of task-evoked oxidative metabolism provide quantitative estimates of $CMRO_2$, which can help understand neuronal differences in oxidative metabolism across age groups even if they have differences in vascular health (Ances et al., 2009; Gauthier et al., 2013; Hutchison et al., 2013a, 2013b; Mohtasib et al., 2012). However, for at least some of this work, the motivation has been to distinguish the physiological components underlying BOLD in attempts to more narrowly isolate the age differences in $CMRO_2$. For instance, Ances and colleagues (2009) found age differences in M, vasodilatory capacity, as the principal difference between their young and old groups. Hutchison and colleagues (2013a), however, isolated the differences to the decoupling change in CBF relative to $CMRO_2$, i.e. decoupling of CBF and $CMRO_2$, implicating neurovascular impairment as an underlying factor of neural efficiency (Neubauer and Fink, 2009; Rypma et al., 2007, 2005). Measuring task-induced relative changes with the calibrated approach relies on several assumptions, including a constant coupling between cerebral blood flow and cerebral blood volume across individuals and brain regions (Gauthier and Fan, 2019). This could lead to ambiguity of interpretation as the baseline may also be changing with age. Yet the CBF-CBV coupling seems to be regionally-specific, and depend on disease stages (Gauthier and Fan, 2019), therefore requiring additional physiologic measures (Germuska and Wise, 2019; Hua et al., 2019). These factors may explain the low frequency of use of such calibrated fMRI.

### 3.1.4  *General remarks on normalisation and calibrating techniques*

Advancing age is associated with multiple alterations in cellular and structural vasculature, leading to multiple physiological changes that directly influence the BOLD signal. However, in contrast to the >10,000 fMRI papers on ageing over the last 20 years, there are fewer than 100 papers addressing calibration and normalisation techniques of fMRI-BOLD signal, and fewer than 20 independent studies of ageing that have applied these techniques. Future studies of ageing should consider the above correction methods, as we expand in Section 4.

A major issue is that nearly all the approaches reviewed above assume that vascular (age-related) factors have a linear influence on BOLD signal. However, there are non-linear influences on the BOLD signal (Mitsis et al., 2004), which is rarely factored in analysis of BOLD data. Individual differences in vascular factors may occur singly, though more often in combination during ageing (Kalaria and Hase, 2019). Moreover, these effects might be relatively independent of one another in early adulthood, but become increasingly coupled with advancing age. It is difficult to define which particular vascular factor might be primarily responsible for age-related changes and the degree and extent of their influence on BOLD signal. This is further complicated by the drugs and medication (e.g. used to normalise blood pressure) that are more often taken in old age. More research is needed to test which correction method (or combination of correction methods) can best correct for cerebrovascular influences to the BOLD signal in ageing studies.

### 3.2  Neuronal integration

The second class of approaches focuses on estimating neuronal contributions to the BOLD signal using independent measures of neural function. The advantages of such an integration approach, as opposed to working solely with measures of neural function, are discussed below. An implicit assumption of these approaches is that they explain variability in neural activity (Figure 2), but not variability related to non-neuronal physiological signals. It is also important to note that such integration will work for age effects detected jointly by both modalities, but neural signals identified uniquely by either modality may remain undetected in the data.

Electroencephalography (EEG) and magnetoencephalography (MEG, together M/EEG) are two widely used non-invasive techniques in neuroscience, and ageing. Although each technique provides important insights in isolation, there are advantages to integrating fMRI and M/EEG in a multimodal approach that is more powerful than each one alone (He et al., 2018). We introduced fMRI as an indirect measure of neural activity with a temporal resolution of seconds, but with a spatial resolution of millimetres. M/EEG, on the other hand, directly measure millisecond electromagnetic activity from large populations of neurons, but at the cost of far worse spatial resolution, particularly for sources of that activity that are deep in the brain. Therefore, combining evidence from M/EEG and fMRI-based techniques can to some extent complement the inherent limitations within each individual imaging modality (Geerligs and Tsvetanov, 2016; He et al., 2018; Sotero and Trujillo-Barreto, 2008). For example, M/EEG can be used to identify neuronal components and events beyond the temporal resolution of fMRI (Laufs, 2008; Ritter and Villringer, 2006), while fMRI can be used to improve the spatial resolution of M/EEG signals (He and Liu, 2008; Henson et al., 2011).

The primary neural source of BOLD signal is synaptic activity in the grey matter, rather than spiking activity, as indicated by a closer relationship of the BOLD signal to local field potentials than multiunit activity recordings (Lauritzen and Gold, 2003; Logothetis et al., 2001; Viswanathan and Freeman, 2007). BOLD-M/EEG associations span multiple frequencies, although those in the gamma band appear most notable. These gamma oscillations (>30hz) are themselves too fast for BOLD to follow, but fluctuations in their power or amplitude envelopes typically fall in similar frequency range as the BOLD signal. Nonetheless, it is important to consider all neuronal frequencies together as a collective account of the fMRI signal, even if differential contributions are found across frequencies (Brookes et al., 2011; Hipp and Siegel, 2015; Scheeringa et al., 2011; Wen and Liu, 2016) and even if different frequencies are differently affected by ageing.

The main advantage MEG has over EEG is that the neuronal sources, though difficult to localise precisely, are better localised than with EEG. Thus while EEG has been used to study neurocognitive ageing for many years, MEG is making an increasing contribution, particularly in combination with formal models of neuronal circuits and their dynamics (Bruffaerts et al., 2019; Lin et al., 2018; Moran et al., 2014; Price et al., 2017; Susi et al., 2019). However, reports that combine task-based MEG with task-based fMRI are only starting to emerge (Bruffaerts et al., 2019), and the combination of MEG with calibrated fMRI in cognitive experiments (Stickland et al., 2019) allows integration of MEG, BOLD and CBF responses to better study differences in neurovascular coupling with ageing.

One advantage that EEG has over MEG however is that it can be acquired simultaneously with fMRI, which is especially important for task-free states that are hard to replicate when EEG and fMRI are run separately (He et al., 2018). EEG has been used to separate neural from vascular components of the BOLD signal (Ritter and Villringer, 2006) and decompose subcomponents of the haemodynamic response function (Mayhew et al., 2010; Mullinger et al., 2013). One challenge for concurrent EEG-fMRI remains the strong magnetic interference in EEG signals (Allen et al., 2000) though this can be mostly overcome during data processing (Liu et al., 2012). In the context of ageing, a study using concurrent EEG-fMRI reported a small number of age-related BOLD components that were associated with EEG (Balsters et al., 2013), suggesting that the combination of both methods can better dissociate neural from non-neural signals than fMRI alone. In addition, there was a set of BOLD components that were not related to EEG components, and vice versa, and it remains unknown whether these components have a neuronal or non-neuronal origin. Since MEG and EEG have different sensitivities to different types of neuronal sources (depending on the orientation and depth of the underlying synaptic currents; Ahlfors et al., 2010; Krishnaswamy et al., 2017), it is advantageous to combine them too (Henson et al., 2011; Muthuraman et al., 2015). Indeed, in future work, EEG data could be acquired simultaneously with MEG and then simultaneously with fMRI, so as to provide a bridge between all three modalities.

There are other techniques, such as measures of glucose utilization (Thompson et al., 2016), beta-amyloid burden (Devous et al., 2012), synaptic density (Chen et al., 2018), optical imaging (Fabiani et al., 2014), PET markers of neuroinflammation (Malpetti et al., 2019; Passamonti et al., 2019), MR spectroscopic measures of the neurotransmitters (Kantarci et al., 2010; Murley and Rowe, 2018) and even non-invasive brain stimulation (Davis et al., 2017; Tan et al., 2019) that may further help understand the basis of BOLD signals, but are beyond the scope of this review.

### 3.3   fMRI modelling and signal decomposition

The third class of approaches focuses on formal modelling or statistical decomposition of the relative contribution of vascular and neuronal factors in the observed BOLD signal. Formal modelling approaches include linear models like the GLM, in which age-related variation can be captured by basis functions, as well as more complex, nonlinear, biophysical models that use differential equations that capture how neuronal events elicit variations in CBF, CBV and CVR and then ultimately changes in the BOLD signal (Buxton et al., 1998; Friston et al., 2000). Fitting these models to BOLD data results in estimates of various parameters that can then be related to age. Statistical decomposition approaches, on the other hand, are data-driven, such as principal component analysis (PCA) and independent component analysis (ICA). Both modelling and decomposition approaches can be useful even when there are no other measures of vascular or neuronal signals (just BOLD data). However, the application and interpretation of these approaches need to be treated with care: biophysical modelling for example typically operates within a high-dimensional space with highly covarying parameters (often requiring prior constraints on the parameters, based on other physiological knowledge), while decomposition techniques will optimise the selection of signals based on a specific statistical criterion; e.g. PCA optimises for variance, while ICA optimises for independence.

*3.3.1 The haemodynamic response function*

The temporal evolution of the BOLD response to a brief burst of neuronal activity (an impulse response) is characterised as the *haemodynamic response function* (HRF). The HRF typically peaks at 5-6s, followed by an undershoot that lasts 20-30s. The precise HRF shape varies across cortical regions, individuals and brain states. This variability is caused by variability in neurovascular coupling and cerebrovascular function, even in the presence of unchanged levels of neural activity (Handwerker et al., 2012; Menon, 2012). One notable finding comes from Cohen and colleagues (Cohen et al., 2002), who demonstrated that varying levels of CBF (induced by hypercapnia, normocapnia and hypocapnia) mediated the onset time, time-to-peak and amplitude of the HRF under the same visual stimulation. The HRF has also been shown to change with genetic, hormonal and other systemic fluctuations (Elbau et al., 2018; Shan et al., 2016; Tong et al., 2019). Regional variability in the HRF is partly dictated by the size of surrounding blood vessels, e.g. regions with larger draining veins have a more delayed HRF (Handwerker et al., 2004; Havlicek and Uludağ, 2020; Taylor et al., 2018). While some fMRI studies use a temporal basis set within the GLM to allow for variations in HRF shape across voxels/individuals (Friston et al., 1998a, 1998b) many use a single, "canonical" HRF. In the latter case, any inferences about age differences in neuronal activity are complicated if there is a systematic age-related bias in the HRF (D'Esposito et al., 1999). For example, a 2-second mis-estimation in the latency of the HRF could (artificially) decrease the magnitude of the estimated neural response by 38% (Handwerker et al., 2004). One potential way to address this would be to demonstrate the specificity of the findings to the condition of interest, but not other contrasts in the experiment.

Several studies have used multiple temporal basis functions with the GLM to capture age-related variations in the HRF shape, including approaches where estimation of the basis function parameters is jointly optimised with estimation of brain activity (Degras and Lindquist, 2014; Pedregosa et al., 2015). However, the results have been mixed (see West et al., 2019), which might reflect variability across studies in the tasks, the use of relatively small sample sizes and biased selection of participants (particularly when older volunteers are more healthy than average). The recent study by West et al. (2019) addressed these problems by using a large, population-derived cohort called Cam-CAN, in which a simple sensorimotor task was optimised for detection of HRF shape (Shafto et al., 2014). This study found extensive effects of age on the HRF, particularly its latency, in many brain regions, despite the fact that there were no performance differences between young and old adults (although latencies of neuronal responses were not directly measured).

*3.3.2 Dynamic causal modelling*

Dynamic Causal Modelling (DCM) is a model-based approach to studying brain connectivity (Friston et al., 2003), which includes a biophysical model of the BOLD response (Buxton et al., 1998; Friston et al., 2000). DCM uses a Bayesian framework to simultaneously estimate parameters capturing neural activity (and connectivity) and parameters capturing the vascular mapping of that activity to the BOLD response. The neural activity can either be defined by experimental manipulations (Friston et al., 2003) or by assumptions about the endogenous fluctuations that occur in task-free states like rest (Razi et al., 2015). Importantly, the simultaneous optimization of neuronal and vascular models (unlike in GLM approaches above) means that differences in the estimated vascular parameters (e.g, due to age) are, in theory, uncontaminated by any differences in neuronal parameters. However, the model can be under-determined (more degrees of freedom in the model than in the data), requiring strong priors on some of the parameters to regularize the models (Daunizeau et al., 2011). These include priors based on physiological data from previous studies, or shrinkage priors that require strong evidence in order for posterior estimates to differ from their prior expectation. The model optimization is made with reference to (log)-model evidence, which accounts for both model accuracy and model complexity.

We applied DCM to resting-state data from 635 adults aged 18-88 in the CamCAN dataset (Tsvetanov et al., 2016). A notable finding was that neural and haemodynamic parameters were independent predictors of age, supporting the hypothesis of separable mechanisms leading to age alterations in neural and vascular function. Furthermore, the neural (connectivity) parameters were related to cognitive ability, and this relationship was moderated by age, demonstrating the behavioural-relevance of this approach to neurocognitive ageing. Interestingly, the same relationship to cognitive ability was not observed with traditional (correlational) analysis of BOLD functional connectivity, which confounds neural and vascular components of the BOLD signal. These findings motivate the use of modelling techniques like DCM to separate neural and vascular components of the BOLD signal.

*3.3.3 Independent Component Analysis (ICA)*

ICA is a data-driven approach to extract signals (components) that are maximally independent across a dimension, such as across space when applied to fMRI images. McKeown and colleagues were some of the first to apply spatial ICA to fMRI BOLD data under the assumption that signal from each voxel represents a linear mixture of source signals (McKeown et al., 1998). Each ICA component consists of a spatial pattern across voxels associated with a common BOLD timecourse. These components are often dominated by neural or non-neural signals (Tong and Frederick, 2014), and their spatial distribution (and/or power spectrum) can sometimes be used to identify vascular components (e.g, around the Circle of Willis) or other noise sources (like motion artifacts, which often appear around the edge of the brain). Another way to separate BOLD from non-BOLD components is to combine ICA with multi-echo fMRI: only ICA components dominated by BOLD signal should show a linear dependency on echo-time (TE) (Kundu et al., 2017)

Most studies use ICA to extract functional networks from task-free fMRI data (Campbell et al., 2015; Tsvetanov et al., 2018, 2016) or structured sources of signal from morphological, vascular and neuroinflammatory measures (Passamonti et al., 2019; Tsvetanov et al., 2015). In task-based BOLD studies, the ICA approach offers multiple advantages over the traditional GLM approach (Xu et al., 2013). It can separate and remove non-neuronal signals for improving the sensitivity of subsequent task-based GLM analysis (Aron and Poldrack, 2006; Tsvetanov et al., 2018). ICA can also identify task-based BOLD changes in a model-free manner that can minimise sensitivity to variation in the HRF shape (LeVan and Gotman, 2009; Schöpf et al., 2011). Finally, ICA can dissociate between multiple concurrent processes associated with common regions under varying cognitive states (Samu et al., 2017). Despite these advantages, future studies need to benchmark the efficiency of ICA to control for age-differences in neurovascular coupling against other data-driven decomposition approaches (Bethlehem et al., 2020) and normalisation methods of task-based BOLD.

## 4 Towards neuro-vascular integration

The previous sections highlighted methodological approaches to addressing age-related changes in cerebrovascular function at the cellular, structural and physiological level in the context of BOLD fMRI and neurocognitive ageing. This methodological emphasis reflects the currently-dominant aim in neurocognitive ageing: to relate cognitive changes to changes in brain activity, such that changes in vascular components are confounds of no interest. Here we argue that, rather than ignoring or correcting for such confounds, we need a better understanding of the neurovascular contribution to neurocognitive ageing, and formally integrate vascular changes into models of successful ageing.

Vascular mechanisms in the brain do not simply control blood flow to support the metabolic needs of neurons, but lead to complex neurovascular interactions that shape neuronal function in health and disease (Abdelkarim et al., 2019; Iadecola, 2017; Wåhlin and Nyberg, 2019). Microvascular changes lead to lacunar infarcts, cortical and subcortical microinfarcts, microbleeds and diffuse white matter disintegration, which involves myelin loss and axonal abnormalities (Wardlaw et al., 2013); all of which potentially impact cognition. Brain areas in relatively sparse regions of the microvascular network, including deeper structures and white matter, are particularly vulnerable, predicting specificity in resulting cognitive deficits (Wåhlin and Nyberg, 2019). Age-related deficits in cognitive function have been linked to cardiovascular risk factors (Baumgart et al., 2015), white matter hyperintensities (Puzo et al., 2019), increased pulsatility (Wåhlin and Nyberg, 2019) and neurovascular coupling impairment (Abdelkarim et al., 2019), which may act through independent pathways (Stefanidis et al., 2019). Furthermore, improvement in cognitive function has been linked to increase in cardiovascular health (Barnes, 2015). These findings suggest that the components of cerebrovascular function are not simply confounders that obscure brain-behaviour relationships, but are synergistic factors that facilitate maintenance and improvement of cognitive function across the lifespan (Abdelkarim et al., 2019; Wåhlin and Nyberg, 2019). Therefore, formal integration of neurovascular knowledge provides an opportunity for a more comprehensive understanding of successful cognitive function in ageing.

Current models of neurovascular ageing (Iadecola, 2017; Kisler et al., 2017) provide an array of biological pathways leading to global brain tissue loss/atrophy and cognitive deficits, mainly in age-related neurodegeneration. However, such models are suboptimal for characterising healthy and successful ageing, where cognitive function is maintained in the presence of brain atrophy (Tsvetanov et al., 2019a). In addition, the link between age-related changes in brain tissue and cognition is surprisingly weak, and it has proven difficult to establish region-by-region correlations between brain structure and cognitive function (Boekel et al., 2015). Moreover, not all cognitive abilities decline with age, nor do all older adults show cognitive decline at the same rate. Studying the effect of neurovascular ageing on brain atrophy or global cognitive decline on its own is insufficient for understanding the complex pattern of cognitive diversity and increasing individual variability in healthy ageing (Shafto et al., 2019; Small et al., 2011).

In the field of neurocognitive ageing, the advent of functional imaging and its early emphasis on functional segregation bolstered the idea that the brain can flexibly respond to age or tissue loss, by recruiting additional brain regions to support cognitive functions (Grady, 2012). Many theories of cognitive ageing have since emerged (Grady, 2012), some proposing that the recruitment of additional brain regions improves performance, while others suggest it can impede performance (Park et al., 2004). Currently, there are three general models of successful ageing in terms of sustained cognitive performance: *maintenance, reorganisation* and *reserve* (Cabeza et al., 2018), which are not necessarily fully compatible. In our view, these models demand more sophisticated interpretation of BOLD fMRI (Morcom and Johnson, 2015) through the integration of neurovascular ageing (Tsvetanov et al., 2016, 2015). It is important to ask whether and how multiple cerebrovascular components (in models of neurovascular ageing) independently and synergistically explain multiple profiles of neural function leading to cognitive diversity in ageing (Tsvetanov et al., 2018).

We propose that one should consider cerebrovascular function as an additional predictor in the modelling of brain-behaviour relationships, rather than simply a normalisation or confounding variable. This will provide a more complete interpretation of the unique and shared contributions to brain-behaviour relationships. For example, the shared variance in task performance explained by vascular and neural signals indicates the presence of a common underlying factor. Conversely, the unique variance explained by neural signals suggests that the effects are beyond differences in cerebrovascular function. Finally, unique variance explained by vascular signals may indicate that the neuronal estimates are insufficient to capture all behavioural variability and an improved definition of the neuronal estimates should be reconsidered. These scenarios are plausible in isolation or in combination with one another, but importantly their consideration provides an empirical motivation to understand what determines cognitive diversity in ageing. Furthermore, modelling and reporting the effects of cerebrovascular function on the brain-behaviour relationship is in the spirit of maximising internal validity (such as confounding) (Lederer et al., 2019), avoiding pitfalls of modular analysis (Lindquist et al., 2019), transparent reporting of results, facilitation of replication and interpretation of findings within the context of the limitations of the research methodology providing the signals of interest.

In summary, we argue for the integration of neurovascular and neurocognitive research on biological, theoretical, methodological and analytical grounds. We propose that future research focuses on the interplay of vascular and neural factors for maintaining mental health across the lifespan (i.e. successful ageing) using a multi-modal, integrative approach. Integration of neurovascular and neurocognitive ageing could provide new insights into the fundamental mechanisms that regulate brain health and mental wellbeing. Importantly, it will determine the extent to which these factors relate to neural function, relate to cognitive performance, and are associated with individual differences in lifestyle, demography, genetics and health. This will provide a bridge between modifiable risk and protective factors, neurovascular function and cognitive ability across the healthy adult lifespan.

## 5    Conclusion

With recent advances in fMRI BOLD imaging, much has been learned about the effects of age on neurovascular and neurocognitive function. It is clear that neurovascular and neuronal signals both contribute to fMRI BOLD signal, and that their interaction affects the interpretations one can draw about neurocognitive ageing. To understand the effect of ageing on brain function, a variety of techniques have been developed and validated that separate vascular from neuronal signals in BOLD-fMRI data. However, only a small fraction of fMRI studies of ageing have adopted such approaches in their analysis. We argue on biological, theoretical and analytical grounds for a better understanding of their relative contributions to functional magnetic resonance imaging (fMRI). Vascular and neuronal contributions can be formally integrated in models of successful ageing, avoiding common misinterpretations of fMRI and complementing the limitations within individual modalities. Only by first understanding these mechanisms and their interactions can we subsequently address a major challenge that pervades neurovascular and neurocognitive ageing: to characterize the effects of healthy and pathological ageing at the level of vascular and neuronal network structures of the human brain across the lifespan.

https://doi.org/10.1002/hbm.20307

Handwerker, D.A., Gonzalez-Castillo, J., D'Esposito, M., Bandettini, P.A., 2012. The continuing challenge of understanding and modeling hemodynamic variation in fMRI. Neuroimage. https://doi.org/10.1016/j.neuroimage.2012.02.015

Handwerker, D.A., Ollinger, J.M., D'Esposito, M., 2004. Variation of BOLD hemodynamic responses across subjects and brain regions and their effects on statistical analyses. Neuroimage 21, 1639–1651. https://doi.org/10.1016/j.neuroimage.2003.11.029

Hare, H. V., Germuska, M., Kelly, M.E., Bulte, D.P., 2013. Comparison of $CO_2$ in air versus carbogen for the measurement of cerebrovascular reactivity with magnetic resonance imaging. J. Cereb. Blood Flow Metab. 33, 1799–1805. https://doi.org/10.1038/jcbfm.2013.131

Hatazawa, J., Shimosegawa, E., Satoh, T., Toyoshima, H., Okudera, T., 1997. Subcortical Hypoperfusion Associated With Asymptomatic White Matter Lesions on Magnetic Resonance Imaging. Stroke 28, 1944–1947. https://doi.org/10.1161/01.STR.28.10.1944

Havlicek, M., Uludağ, K., 2020. A dynamical model of the laminar BOLD response. Neuroimage 204. https://doi.org/10.1016/j.neuroimage.2019.116209

He, B., Liu, Z., 2008. Multimodal Functional Neuroimaging: Integrating Functional MRI and EEG/MEG. IEEE Rev. Biomed. Eng. 1, 23–40. https://doi.org/10.1109/RBME.2008.2008233

He, B., Sohrabpour, A., Brown, E., Liu, Z., 2018. Electrophysiological Source Imaging: A Noninvasive Window to Brain Dynamics. Annu. Rev. Biomed. Eng. 20, 171–196. https://doi.org/10.1146/annurev-bioeng-062117-120853

Heijtel, D.F.R., Mutsaerts, H.J.M.M., Bakker, E., Schober, P., Stevens, M.F., Petersen, E.T., van Berckel, B.N.M., Majoie, C.B.L.M., Booij, J., van Osch, M.J.P., vanBavel, E., Boellaard, R., Lammertsma, A.A., Nederveen, A.J., 2014. Accuracy and precision of pseudo-continuous arterial spin labeling perfusion during baseline and hypercapnia: A head-to-head comparison with 15O H2O positron emission tomography. Neuroimage 92, 182–192. https://doi.org/10.1016/J.NEUROIMAGE.2014.02.011

Henson, R.N., Flandin, G., Friston, K.J., Mattout, J., 2010. A Parametric Empirical Bayesian framework for fMRI-constrained MEG/EEG source reconstruction. Hum. Brain Mapp. 31, 1512–1531. https://doi.org/10.1002/hbm.20956

Henson, R.N., Mattout, J., Phillips, C., Friston, K.J., 2009. Selecting forward models for MEG source-reconstruction using model-evidence. Neuroimage 46, 168–176.38

# 7  Acknowledgements and funding

We thank Wiktor Olszowy for valuable comments. This work is supported by the British Academy (PF160048), the Guarantors of Brain (G101149), the Wellcome Trust (103838), the Medical Research Council (SUAG/051 G101400; and SUAG/046 G101400), European Union's Horizon 2020 (732592) and the Cambridge NIHR Biomedical Research Centre.